\documentclass[aps,prd,floatfix,groupedaddress,nofootinbib,superscriptaddress,twocolumn]{revtex4-1}

\usepackage{amsmath,amssymb,bm,color,float,graphicx,mathrsfs}
\usepackage[hidelinks]{hyperref}
\hypersetup{colorlinks=true,linkcolor=blue,citecolor=blue,urlcolor=blue}
\usepackage{Macro}

\def\a{\alpha}
\def\ACB{A_{\rm CB}}

\def\esta{\widehat{\alpha}}
\def\uesta{\ol{\alpha}}

\def\hX{\widehat{X}}
\def\hCXX{\widehat{C}^{\rm XX}}
\def\hCaa{\widehat{C}^{\a\a}}
\def\Caa{C^{\a\a}}

\def\hA{\widehat{A}}
\def\sign{{\rm sign}}
\def\actpol{ACTPol}

\def\gangle{0.12\pm 0.06}
\def\ACBconst{0.10} 
\def\Hgconst{4.0\times 10^{-2}}
\def\ptereal{0.99}
\def\pterealLminA{0.85}
\def\pterealLminB{0.94}
\def\pterealbinnA{0.77}
\def\pterealbinnB{0.88}
\def\lmin{200}
\def\areadeep{456}
\def\areaboss{1633}

\begin{document}

\title{The Atacama Cosmology Telescope: Constraints on Cosmic Birefringence}

\author{Toshiya Namikawa$^*$}
\affiliation{Department of Applied Mathematics and Theoretical Physics, University of Cambridge, Wilberforce Road, Cambridge CB3 0WA, United Kingdom}
\author{Yilun Guan}
\affiliation{Department of Physics and Astronomy, University of Pittsburgh, Pittsburgh, PA 15260, USA}
\author{Omar Darwish}
\affiliation{Department of Applied Mathematics and Theoretical Physics, University of Cambridge, Wilberforce Road, Cambridge CB3 0WA, United Kingdom}
\author{Blake D. Sherwin}
\affiliation{Department of Applied Mathematics and Theoretical Physics, University of Cambridge, Wilberforce Road, Cambridge CB3 0WA, United Kingdom}

\author{Simone Aiola}
\affiliation{Center for Computational Astrophysics, Flatiron Institute, 162 5th Avenue, New York, NY, USA 10010}

\author{Nicholas Battaglia}
\affiliation{Department of Astronomy, Cornell University, Ithaca, New York 14853, USA}

\author{James A. Beall}
\affiliation{National Institute of Standards and Technology, 325 Broadway, Boulder CO, 80305}

\author{Daniel T. Becker}
\affiliation{National Institute of Standards and Technology, 325 Broadway, Boulder CO, 80305}

\author{J. Richard Bond}
\affiliation{Canadian Institute for Theoretical Astrophysics, University of Toronto, 60 St. George St., Toronto, ON M5S 3H8, Canada}

\author{Erminia Calabrese}
\affiliation{School of Physics and Astronomy, Cardiff University, Cardiff CF10 3AT, United Kingdom}

\author{Grace E. Chesmore}
\affiliation{Department of Physics, University of Michigan, Ann Arbor, MI 48109}

\author{Steve K. Choi}
\affiliation{Department of Astronomy, Cornell University, Ithaca, New York 14853, USA}

\author{Mark J. Devlin}
\affiliation{Department of Physics and Astronomy, University of Pennsylvania, 209 South 33rd Street, Philadelphia, PA 19104}

\author{Joanna Dunkley}
\affiliation{Department of Astrophysical Sciences, Princeton University, 4 Ivy Lane, Princeton, NJ, USA 08544}
\affiliation{Joseph Henry Laboratories of Physics, Jadwin Hall, Princeton University, Princeton, NJ 08544, USA}

\author{Rolando D\"unner}
\affiliation{Instituto de Astrof\'isica and Centro de Astro-Ingenier\'ia, Facultad de F\'isica, Pontificia Universidad Cat\'olica de Chile, Av. Vicu\~na Mackenna 4860, 7820436, Macul, Santiago, Chile}

\author{Anna E. Fox}
\affiliation{National Institute of Standards and Technology, 325 Broadway, Boulder CO, 80305}

\author{Patricio A. Gallardo}
\affiliation{Department of Physics, Cornell University, Ithaca, NY14853, USA}

\author{Vera Gluscevic}
\affiliation{Department of Physics and Astronomy University of Southern California Los Angeles, CA 90089-0484}

\author{Dongwon Han}
\affiliation{Physics and Astronomy Department, Stony Brook University, Stony Brook, NY 11794, USA}
\affiliation{Center for Computational Astrophysics, Flatiron Institute, 162 5th Avenue, New York, NY, USA 10010}

\author{Matthew Hasselfield}
\affiliation{Center for Computational Astrophysics, Flatiron Institute, 162 5th Avenue, New York, NY, USA 10010}

\author{Gene C. Hilton}
\affiliation{National Institute of Standards and Technology, 325 Broadway, Boulder CO, 80305}

\author{Adam D. Hincks}
\affiliation{Canadian Institute for Theoretical Astrophysics, University of Toronto, 60 St. George St., Toronto, ON M5S 3H8, Canada}

\author{Ren\'ee Hlo\v{z}ek}
\affiliation{Dunlap Institute for Astronomy and Astrophysics, University of Toronto, 50 St George Street, Toronto ON, M5S 3H4, Canada}
\affiliation{David A. Dunlap Department of Astronomy and Astrophysics, University of Toronto, 50 St George Street, Toronto ON, M5S 3H4, Canada}

\author{Johannes Hubmayr}
\affiliation{Affiliation: Quantum Sensors Group, NIST, Boulder, 325 Broadway, Boulder, CO 80305, USA}

\author{Kevin Huffenberger}
\affiliation{Department of Physics, Florida State University 609 Keen Physics Building Tallahassee, FL, 32306, USA}

\author{John P. Hughes}
\affiliation{Department of Physics and Astronomy, Rutgers University, 136 Frelinghuysen Road, Piscataway, NJ 08854-8019, USA}

\author{Brian J. Koopman}
\affiliation{Department of Physics, Yale University, New Haven, CT 06511, USA}

\author{Arthur Kosowsky}
\affiliation{Department of Physics and Astronomy, University of Pittsburgh, Pittsburgh, PA 15260, USA}

\author{Thibaut Louis}
\affiliation{Université Paris-Saclay, CNRS/IN2P3, IJCLab, 91405 Orsay, France}

\author{Marius Lungu}
\affiliation{Joseph Henry Laboratories of Physics, Jadwin Hall, Princeton University, Princeton, NJ 08544, USA}

\author{Amanda MacInnis}
\affiliation{Physics and Astronomy Department, Stony Brook University, Stony Brook, NY 11794, USA}

\author{Mathew S. Madhavacheril}
\affiliation{Centre for the Universe, Perimeter Institute for Theoreticaxl Physics, Waterloo, ON, Canada N2L 2Y5}
\affiliation{Department of Astrophysical Sciences, Princeton University, 4 Ivy Lane, Princeton, NJ, USA 08544}

\author{Maya Mallaby-Kay}
\affiliation{Department of Physics, University of Michigan, Ann Arbor, MI 48109}

\author{Lo\"ic Maurin}
\affiliation{Universit\'e Paris-Saclay, CNRS, Institut d'astrophysique spatiale, 91405, Orsay, France.}

\author{Jeffrey McMahon}
\affiliation{Kavli Institute for Cosmological Physics, University of Chicago, 5640 S. Ellis Ave., Chicago, IL 60637, USA}
\affiliation{Department of Astronomy and Astrophysics, University of Chicago, 5640 S. Ellis Ave., Chicago, IL 60637, USA}
\affiliation{Department of Physics, University of Chicago, Chicago, IL 60637, USA}
\affiliation{Enrico Fermi Institute, University of Chicago, Chicago, IL 60637, USA}

\author{Kavilan Moodley}
\affiliation{Astrophysics and Cosmology Research Unit, School of Mathematics, Statistics \& Computer Science, University of KwaZulu-Natal, Durban, 4041, South Africa}

\author{Sigurd Naess}
\affiliation{Center for Computational Astrophysics, Flatiron Institute, 162 5th Avenue, New York, NY, USA 10010}

\author{Federico Nati}
\affiliation{Department of Physics, University of Milano - Bicocca, Piazza della Scienza, 3 - 20126, Milano (MI), Italy}

\author{Laura B. Newburgh}
\affiliation{Department of Physics, Yale University, New Haven, CT 06520,
USA}

\author{John P. Nibarger}
\affiliation{National Institute of Standards and Technology, 325 Broadway, Boulder CO, 80305}

\author{Michael D. Niemack}
\affiliation{Department of Physics, Cornell University, 109 Clark Hall, Ithaca, NY 14853, USA}
\affiliation{Department of Astronomy, Cornell University, Ithaca, NY 14853, USA}

\author{Lyman A. Page}
\affiliation{Joseph Henry Laboratories of Physics, Jadwin Hall, Princeton University, Princeton, NJ 08544, USA}

\author{Frank J. Qu}
\affiliation{Department of Applied Mathematics and Theoretical Physics, University of Cambridge, Wilberforce Road, Cambridge CB3 0WA, United Kingdom}

\author{Naomi Robertson}
\affiliation{Institute of Astronomy, University of Cambridge, Madingley Rd, Cambridge CB3 0HA}

\author{Alessandro Schillaci}
\affiliation{Department of Physics, California Institute of Technology, Pasadena, California 91125, USA}

\author{Neelima Sehgal}
\affiliation{Physics and Astronomy Department, Stony Brook University, Stony Brook, NY 11794, USA}
\affiliation{Center for Computational Astrophysics, Flatiron Institute, 162 5th Avenue, New York, NY, USA 10010}

\author{Crist\'obal Sif\'on}
\affiliation{Instituto de F\'isica, Pontificia Universidad Cat\'olica de Valpara\'iso, Casilla 4059, Valpara\'iso, Chile}

\author{Sara M. Simon}
\affiliation{Department of Physics, University of Michigan, 450 Church St., Ann Arbor, MI 48109}

\author{David N. Spergel}
\affiliation{Department of Astrophysical Sciences, Princeton University, 4 Ivy Lane, Princeton, NJ, USA 08544}
\affiliation{Center for Computational Astrophysics, Flatiron Institute, 162 5th Avenue, New York, NY, USA 10010}

\author{Suzanne T. Staggs}
\affiliation{Joseph Henry Laboratories of Physics, Jadwin Hall, Princeton University, Princeton, NJ 08544, USA}

\author{Emilie R. Storer}
\affiliation{Joseph Henry Laboratories of Physics, Jadwin Hall, Princeton University, Princeton, NJ 08544, USA}

\author{Alexander van Engelen}
\affiliation{School of Earth and Space Exploration, Arizona State University, Tempe, AZ, 85287, USA}

\author{Jeff van Lanen}
\affiliation{National Institute of Standards and Technology, 325 Broadway, Boulder CO, 80305}

\author{Edward J. Wollack}
\affiliation{NASA Goddard Spaceflight Center, 8800 Greenbelt Road, Greenbelt, Maryland 20771, USA}


\renewcommand{\thefootnote}{\fnsymbol{footnote}}
\footnote[0]{$^*$ Corresponding author: 
\href{tn334@cam.ac.uk}{tn334@cam.ac.uk}}

\date{\today}

\begin{abstract}
We present new constraints on anisotropic birefringence of the cosmic microwave background polarization using two seasons of data from the Atacama Cosmology Telescope covering $\areadeep$ square degrees of sky. The birefringence power spectrum, measured using a curved-sky quadratic estimator, is consistent with zero. Our results provide the tightest current constraint on birefringence over a range of angular scales between $5$ arcminutes and $9$ degrees. We improve previous upper limits on the amplitude of a scale-invariant birefringence power spectrum by a factor of between $2$ and $3$. Assuming a nearly-massless axion field during inflation, our result is equivalent to a $2\,\sigma$ upper limit on the Chern-Simons coupling constant between axions and photons of $g_{\alpha\gamma} < \Hgconst{}/H_I$ where $H_I$ is the inflationary Hubble scale. 
\end{abstract} 


\maketitle


\section{Introduction} \label{sec:intro}

The Atacama Cosmology Telescope (ACT) experiment, a 6-m diameter mm-band telescope located in the Atacama Desert in Chile, has completed several seasons of cosmic microwave background (CMB) polarization observations \cite{Thornton:2016:ACT}. These observations have been used to derive a variety of scientific results, for example, via measurements of the CMB power spectrum \cite{ACT:Louis:2016} and gravitational lensing by large-scale structure \cite{ACT16:phi,ACT:phixcib}. Beyond these observables, ACT's CMB polarization data can be used to test for new physics by searching for a rotation of linear polarization as the CMB photons propagate to us from the surface of last scattering. This phenomenon, which is absent in the Standard Model, is referred to as cosmic birefringence.

Several types of beyond-the-Standard-Model physics can source cosmic birefringence. In particular, birefringence of CMB photons can be generated by axion-like particles within a mass range of $10^{-33}\alt m_a\alt 10^{-28}$ eV that couple to photons through a so-called Chern-Simons term (see e.g., \cite{Carroll:1998,Li:2008,Pospelov:2009,Finelli:2009,Hlozek:2017:axion} and a review, \cite{Marsh:2016}). \footnote{Axion-like particles within a mass range of $10^{-22}\alt m_a\alt 10^{-18}$ eV also introduce a time variation of the polarization angle rotation whose oscillation period is from hours to years, and can be tightly constrained by current and future CMB experiments as discussed in \cite{Fedderke:2019:biref}.} The existence of such axion-like particles is a generic prediction of string theory. In addition, birefringence-inducing pseudo-scalar fields could be candidates for an early dark energy mechanism to resolve the current Hubble parameter tension \cite{Capparelli:2019:CB}. Cosmic birefringence can be used as a probe of, e.g., the axion string network \cite{Agrawal:2019:biref}, axion dark matter \cite{Liu:2016dcg}, and also more general Lorentz-violating physics in the context of Standard Model extensions \cite{Leon:2017}. Finally, cosmic birefringence can also be generated by primordial magnetic fields (PMFs) through Faraday rotation of the CMB polarization (e.g., \cite{Kosowsky:1996,Harari:1997,Kosowsky:2004:FR,Yadav:2012b,De:2013,Pogosian:2014}).
The PMF-induced cosmic birefringence has frequency dependence and can be distinguished from that induced by axion-like particles \cite{Gubitosi:2013}.
If the source of the cosmic birefringence is spatially varying, the polarization rotation will be anisotropic (i.e., have different values in different directions in the sky); indeed, anisotropies in the cosmic birefringence are produced naturally by many of the types of beyond-the-Standard-Model physics listed previously (see, e.g., \cite{Carroll:1998,Lue:1999,Li:2006pu,Caldwell:2011,Lee:2015,Leon:2017}).
For example, quintessence models predict both isotropic and anisotropic cosmic birefringence \cite{Caldwell:2011}. In addition, the cosmic birefringence induced by some massless scalar fields does not necessarily produce isotropic cosmic birefringence, and a measurement of the anisotropic birefringence is crucial to constraining such scenarios \cite{Gluscevic:2012qv}. Measurements of, or tight constraints on, the relevant pseudo-scalar fields and other phenomena can hence provide valuable insights into fundamental physics. 

Both isotropic and anisotropic cosmic birefringence have been constrained by several CMB experiments, although the observational effects on the CMB and the methodology to measure these two types of birefringence are different. 
The presence of isotropic cosmic birefringence can be detected in CMB observations because the polarization rotation transfers part of the CMB $E$ mode polarization to $B$ modes and thus creates non-zero odd-parity $EB$ power spectra. Such odd-parity $EB$ power spectra are zero in the standard cosmological model and have hence been used for constraining isotropic cosmic birefringence \cite{Feng:2006:CPT,Gruppuso:2012,Li:2014,Mei:2014iaa}. However, odd-parity power spectra have systematic uncertainties from the global polarization angle calibration \cite{Pagano:2009:biref,Miller:2009:biref,Keating:2013,Hinshaw:2013,P16:rot}; in fact, odd-parity spectra are often used to calibrate the global polarization angle rather than for measuring cosmological signals \cite{B1rot,BKX}. Galactic foreground components in the observed odd-parity spectra can be used to partially break degeneracies between the global polarization angle error and cosmic birefringence effects \cite{Minami:2019:rot}, although the signal-to-noise can decrease somewhat in this process. 

On the other hand, if the cosmic birefringence is anisotropic, we can measure it not only using CMB power spectra \cite{Pospelov:2009} but also by using the fact that the $EB$ correlation varies with direction, which is characteristic of statistical anisotropy \cite{Kamionkowski:2009:derot}. Variations of the polarization rotation angle on angular scale $L$ will mix together $E$ and $B$ modes of different scales, leading to non-zero expectation values in the off-diagonal ($\l\ne\l'$) elements of the CMB covariance [see \eq{Eq:weight} below]. \footnote{In this paper, we use $L$ to denote the multipole of the reconstructed rotation angle and $\l$ to denote the multipole of CMB anisotropies.} We can, therefore, reconstruct the anisotropies of the cosmic birefringence by measuring these off-diagonal correlations, in a manner similar to CMB lensing reconstruction \cite{OkamotoHu:quad}. Other pairs of CMB anisotropies such as temperature and $B$-modes are also correlated, but such correlations generally give lower signal-to-noise ratios for reconstructing birefringence \cite{Yadav:2012a}. We will therefore focus on birefringence reconstruction from $EB$ correlations in this paper. 
Note that the cosmic birefringence and lensing can be estimated separately by using their distinct effects on polarization maps in terms of parity \cite{Kamionkowski:2009:derot}; birefringence introduces rotations with determinate directions and the resulting map is not parity-symmetric, whereas lensing arising from the scalar density field has even parity (see Sec.~\ref{sec:method} for details). 

Multiple publications have presented constraints on anisotropies of the cosmic birefringence using reconstruction methods; these have made use of the WMAP temperature and $B$-modes \cite{Gluscevic:2012qv}, or the polarization data of the POLARBEAR \cite{PB15:rot}, BICEP2/Keck Array \cite{BKIX}, and Planck \cite{Contreras:2017} experiments. The use of the reconstructed cosmic birefringence power spectrum is the most powerful current method for measuring the anisotropies of the cosmic birefringence and indeed gives the best current constraints \cite{BKIX,Contreras:2017}. However, we note that several other publications \cite{Gubitosi:2011:biref,Li:2013,Alighieri:2014yoa,Li:2014,Mei:2014iaa,Liu:2016dcg,Zhai:2019god} also place constraints on anisotropic birefringence by analyzing CMB polarization power spectra.

In this paper, we reconstruct the cosmic birefringence (rotation) field from the CMB polarization using a quadratic estimator analogous to those commonly used in measuring the cosmic deflection field due to gravitational lensing. The estimator includes the effect of sky curvature, which will become increasingly important as low-noise and high-resolution polarization maps extend over larger sky regions. We focus on frequency-independent cosmic birefringence and apply this estimator to data from the Atacama Cosmology Telescope Polarimeter (\actpol), finding a rotation field consistent with zero within measurement uncertainties. Our limits on polarization rotation are the strongest to date over a wide range of scales. 

In Sec.~\ref{sec:datasim}, we describe our data and simulations for the cosmic birefringence reconstruction. In Sec.~\ref{sec:method}, we explain our reconstruction methodology, and in Sec.~\ref{sec:test} we explore potential systematic errors that are relevant for the cosmic birefringence analysis. Sec.~\ref{sec:results} shows our results for the reconstructed spectrum and the resulting constraint on the scale-invariant birefringence spectrum. We discuss implications for axion-like particles in Sec.~\ref{sec:discussion}.

\section{Data and simulations} \label{sec:datasim}

We analyze \actpol\ nighttime polarization data collected from two seasons of observations taken in 2014 and 2015. These data are described in \cite{ACT:Louis:2016}. In this paper, the constraints on cosmic birefringence anisotropies are derived using data from one region of the sky, which we label {\tt D56}. {\tt D56} spans $\areadeep$ deg$^2$ of the sky with the aspect ratio of 1:4 observed in both the 2014 and 2015 seasons at $150$ GHz, and in the 2015 season at $90$ GHz \cite{ACT:Louis:2016,ACT:Aiola:2019,ACT:Choi:2019}; the map has an effective noise level of $14\,\mu$K-arcmin for polarization. In addition to {\tt D56}, we also use another region of the sky for a (swap-patch) null test: the region is called {\tt BOSS-N}; this field was observed during the 2015 season and covers $\areaboss$ deg$^2$ of the sky with an effective noise level of roughly $30\,\mu$K-arcmin in polarization. 
Since the statistical error of the reconstructed cosmic-birefringence spectrum from the {\tt BOSS-N} region is roughly $4$-$5$ times larger than that from {\tt D56}, the improvement of the cosmic birefringence constraint by adding {\tt BOSS-N} is roughly $\lesssim 3$\%. \footnote{The noise spectrum of the reconstructed cosmic-birefringence fields scales as fourth power of the CMB map noise if the CMB maps are noise dominant. In {\tt BOSS-N}, even $EE$ spectrum is not signal-dominant, and the reconstructed noise spectrum is roughly an order of magnitude larger than that of {\tt D56}. Taking into account the sky coverage, the statistical error of the reconstructed spectrum is $4$-$5$ times larger than that of {\tt D56}. The expected improvement on the signal-to-noise of the birefringence spectrum is thus negligible, $\sqrt{1+(1/4)^2}\sim 3\%$.}
Thus, we use the {\tt BOSS-N} data only for a null test. For each region ({\tt D56} and {\tt BOSS-N}), the Fourier-space combined $E$ and $B$ maps are produced from the maps in each frequency, each detector array, and season \cite{ACT:Omar:2020}. 

We use Monte Carlo simulations for the standard $\Lambda$CDM cosmology \footnote{The standard cosmology in this paper is the flat $\Lambda$CDM model parametrized by the six cosmological parameters with values close to the best-fit 2015 Planck parameters \cite{P15:compsep:cmb}.} to test our pipeline, compute the biases in the power spectrum measurement, perform null tests, and calculate the covariance matrix for the $\chi^2$ PTE and likelihood. The simulation includes lensing and realistic effects to mimic the data such as beams and inhomogeneous noise (see \cite{ACT:Louis:2016,ACT:Choi:2019} for the details). Hereafter, we call this simulation the standard simulation. 
Additional simulations including scale-invariant rotation anisotropies with varying amplitudes $\ACB$ (see Sec.~\ref{sec:results} for the definition) are used for pipeline tests and to compute the transfer function for the reconstructed spectrum. To assess the impact of the global polarization angle error on our measurements, we also generate a simulation with an offset in the global polarization angle (see Sec.~\ref{sec:test}). We obtain a candidate dust map by appropriately scaling the Galactic dust simulation of \cite{Vansyngel:2016fbn}, which provides a non-Gaussian full-sky dust $Q/U$ map at $353$~GHz.

\section{Analysis} \label{sec:method}

The rotation angle field, $\alpha(\hatn)$, can be reconstructed from the off-diagonal mode-mode covariance within and between, the $E$ and $B$ modes \cite{Kamionkowski:2009:derot}. An estimator of $\a(\hatn)$ has a quadratic form similar to the lensing estimator. The power spectrum of the anisotropic rotation angle $C_L^{\a\a}$ can be obtained by squaring the rotation estimator and subtracting relevant biases.
We use the curved-sky quadratic estimator to extract the large-scale birefringence anisotropies which are important to constrain the scale-invariant spectrum described later. 
Verification of the method to measure the cosmic birefringence spectrum applied in this paper is described in \cite{Namikawa:2016:rotsim} for a flat-sky analysis. In the extension described here, the estimator in a full-sky formalism employs spherical harmonic transformations instead of Fourier transforms. \footnote{The code used for reconstructing the cosmic birefringence in fullsky is based on \url{https://toshiyan.github.io/clpdoc/html/}.} 

In order to account for ground and atmospheric noise, we begin by filtering out the Fourier modes $|\l_x|<90$ and $|\l_y|<50$ of the $E$ and $B$ maps produced by combining seasons, frequencies and arrays in Fourier space \cite{ACT:Omar:2020}. This is the same filter as is applied for the CMB power spectrum and lensing reconstruction analysis \cite{ACT:Louis:2016,ACT16:phi}. This process is performed using a flat-sky Fourier transform. After transforming back to position space, we assign the filtered $E$ and $B$ maps to the {\tt Healpix} grids and compute the harmonic coefficients of the $E$ and $B$ modes. 
Note that the polarization maps are provided at each patch which take into account the curved-sky geometry. The harmonic coefficients computed from these maps thus do not have any distortion due to ignoring the curved-sky geometry. In fact, if such distortion is significant, we need additional filtering process in computing $\ol{X}_{\l m}$, or additional correction to the estimator normalization. As we discussed below, however, we found that the mismatch between the input birefringence spectrum and cross-spectrum between input and reconstructed birefringence fields is very small, and the above distortion is negligible.

The presence of the cosmic birefringence effect rotates the primordial Stokes parameters as \cite{Kamionkowski:2009:derot,Gluscevic:2012qv} 
\al{
    Q'(\hatn)\pm\iu U'(\hatn) = [Q(\hatn)\pm\iu U(\hatn)]\E^{\pm 2\iu \alpha(\hatn)}
    \,. \label{Eq:qurot}
}
Consequently, the rotation angle modifies the CMB $E$ and $B$ modes. The $E$ and $B$ modes are obtained by transforming $Q$ and $U$ maps with the spin-2 spherical harmonics, $Y_{\l m}^{\pm 2}$, as \cite{Kamionkowski:1996:eb,Zaldarriaga:1996:EBdef}: 
\al{
    E_{\l m}\pm \iu B_{\l m} = -\Int{2}{\hatn}{}(Y^{\pm 2}_{\l m})^*[Q(\hatn)\pm\iu U(\hatn)]
    \,. 
}
Thus, the $E$ and $B$ modes in the presence of an anisotropic rotation angle are derived by substituting \eq{Eq:qurot} into the above equation, and are given up to linear order in $\alpha$ by  \cite{Gluscevic:2009,Kamionkowski:2009:derot}: 
\begin{widetext}
\al{
	E'_{\l m}\pm\iu B'_{\l m} 
	= E_{\l m}\pm\iu B_{\l m} + \sum_{LM\l'm'}(-1)^m\Wjm{\l}{L}{\l'}{-m}{M}{m'} W^\pm_{\l L\l'} [E_{\l'm'}\pm\iu B_{\l'm'}] \alpha_{LM} 
	\,, 
}
with
\al{
	W^\pm_{\l_1\l_2\l_3} = \pm 2\zeta^\mp p^\mp_{\l_1\l_2\l_3}\sqrt{\frac{(2\l_1+1)(2\l_2+1)(2\l_3+1)}{4\pi}}\Wjm{\l_1}{\l_2}{\l_3}{-2}{0}{2} 
	\,. 
}
\end{widetext}
Here, $\zeta^+=1$, $\zeta^-=\iu$, and  $p^\pm_{\l_1\l_2\l_3}=[1\pm(-1)^{\l_1+\l_2+\l_3}]/2$ is a parity indicator. The ending parentheses denote the Wigner $3j$-symbol. The off-diagonal elements of the covariance induced by the anisotropies of the rotation angle are given by \cite{Gluscevic:2009}
\al{
	\ave{E'_{\l m}B'_{\l'm'}}\rom{CMB} 
		&= \sum_{LM}\Wjm{\l}{\l'}{L}{m}{m'}{M} f^\alpha_{\l L \l'} \alpha^*_{LM} 
	\,, \label{Eq:weight} 
}
where $\l\not=\l'$, $m\not=-m'$, and the operator $\ave{\cdots}\rom{CMB}$ denotes an ensemble average over the realizations of CMB and noise with a fixed realization of $\a(\hatn)$. The weight function is 
\al{
	f^\alpha_{\l L \l'} &= -W^-_{\l'L\l}\tCEE_{\l}\,, \label{Eq:weight:a}
}
where $\tCEE_\l$ is the lensed $E$-mode power spectrum. The term originating from the lensing $B$ mode is ignored since the improvement of the sensitivity to the polarization rotation anisotropies by the inclusion of this term is negligible \cite{PB15:rot}. Similar to the lensing reconstruction, the unnormalized quadratic estimator of $\a$ is constructed as a convolution of the $E$ and $B$ modes with the weight function of \eq{Eq:weight:a} \cite{Gluscevic:2009}: 
\al{
	\uesta^*_{LM} = \sum_{\l\l'mm'}\Wjm{\l}{\l'}{L}{m}{m'}{M}f^\alpha_{\l L \l'}\ol{E}_{\l m}\ol{B}_{\l'm'}
	\,. \label{Eq:uest}
}
Here, $\ol{E}_{\l m}$ and $\ol{B}_{\l m}$ are the observed multipoles filtered by their inverse variance. We use diagonal filtering, $\ol{X}_{\l m}=\hX_{\l m}/\hCXX_\l$, where $X$ is either $E$ or $B$ and $\hCXX_\l$ is the power spectrum of the observed multipoles, $\hX_{\l m}$. The CMB multipoles at $\lmin\leq\l\leq2048$ are used for our baseline reconstruction, although we also perform the reconstruction for other multipole ranges as a test of the analysis in Sec.~\ref{sec:test}. Finally, we correct for the mean-field bias, $\ave{\uesta_{LM}}$, and normalize to obtain the rotation angle: 
\al{
    \esta_{LM} = A_L(\uesta_{LM}-\ave{\uesta_{LM}}) 
    \,. \label{Eq:aest}
}
The normalization $A_L$ is given by
\al{
	A_L = \frac{1}{2L+1}\sum_{\l\l'}\frac{(f^\alpha_{\l L\l'})^2}{\hCEE_\l\hCBB_{\l'}}
	\,. \label{Eq:Rec:N0}
}
We compute \eqs{Eq:uest,Eq:Rec:N0} with a computationally efficient method as described in Appendix \ref{app:code}. The mean-field bias could be non-zero due to, for example,  survey boundary and beam asymmetry effects \cite{Namikawa:2012:bhe} (see also Appendix \ref{app:mean}). We evaluate the mean-field bias by averaging over the standard simulations, finding that the bias is less than $0.5\%$ of the $1\,\sigma$ statistical error of the cosmic birefringence spectrum over the scales relevant to our analysis. Mean-field bias is also induced by the global polarization angle error, which is not included in the standard simulations. We evaluate this effect in Sec.~\ref{sec:test} and discuss its origin in Appendix \ref{app:mean}.

We note that the quadratic estimator for the cosmic birefringence anisotropies given above is the same as that for the CMB lensing potential, $\phi$, but with a different weight function. In the case of CMB lensing, the off-diagonal covariance of \eq{Eq:weight} is given by \cite{OkamotoHu:quad}
\al{
	\ave{\widetilde{E}_{\l m}\widetilde{B}_{\l'm'}}\rom{CMB} 
		&= \sum_{LM}\Wjm{\l}{\l'}{L}{m}{m'}{M} f^\phi_{\l L \l'} \phi^*_{LM} 
	\,,  
}
where $\widetilde{E}$ and $\widetilde{B}$ are the lensed $E$ and $B$ modes, respectively, and $f^\phi$ is defined as \cite{OkamotoHu:quad}
\al{
	&f^\phi_{\l L \l'} = - \iu p^-_{\l L\l'}\sqrt{\frac{(2\l+1)(2L+1)(2\l'+1)}{16\pi}} 
	\notag \\ 
	&\quad \times [L(L+1)+\l(\l+1)-\l'(\l'+1)] \Wjm{\l'}{L}{\l}{2}{0}{-2} \tCEE_{\l}
	\,. \label{Eq:weight:phi}
}
Here, $p^-_{\l+L+\l'}=[1-(-1)^{\l+L+\l'}]/2$ is the parity indicator. The estimator for the CMB lensing potential is then obtained in the same form as \eq{Eq:uest} but replacing $f$ with $f^\phi$. The normalization is also obtained in the same way. 
The main difference between the properties of $f^\alpha$ and $f^\phi$ is that $f^\alpha$ and $f^\phi$ are only non-zero when $\l+L+\l'$ is even for $f^\alpha$ and odd for $f^\phi$, respectively \cite{Kamionkowski:2009:derot}. This property comes from the difference of the parity symmetry between the lensing potential and cosmic birefringence anisotropies. With this property of $f^\alpha$ and $f^\phi$, our standard estimators can completely separate the cosmic birefringence and lensing potential contributions to the off-diagonal elements of the $EB$ correlation similar to the lensing potential and curl mode decomposition \cite{Namikawa:2011:curlrec}. 

From the reconstructed $\a$, the cosmic birefringence spectrum is also estimated in the same way as the CMB lensing power spectrum \cite{ACT16:phi,P18:phi}, but with a few modifications. The power spectrum of the estimator defined in \eq{Eq:aest} is a four-point function and has a disconnected (or Gaussian) bias, $N_L^0$, which is simply due to the original, unrotated CMB anisotropies and is non-zero even in the absence of birefringence \cite{OkamotoHu:quad,P13:phi,Namikawa:2016:rotsim}. We construct an estimator of the disconnected bias, $\widehat{N_L^0}$, using a realization-dependent algorithm \cite{Namikawa:2012:bhe} and subtract it from the power spectrum of the estimator to extract the cosmic birefringence spectrum. For simulations, for convenience we subtract the ensemble average, $\ave{\widehat{N_L^0}}$, instead of $\widehat{N_L^0}$, which makes less than $1\%$ difference in the $\chi^2$ PTE of the reconstructed cosmic birefringence spectrum and the constraints on the scale-invariant spectrum. Using a realization-dependent bias subtraction makes our measurement of the birefringence spectrum robust to possible mismatches between simulation and data. 
The lenisng bias and N1 bias as shown in \cite{Namikawa:2016:rotsim} are evaluated using simulation from the standard and non-zero birefringence simulations, respectively. The lensing bias is subtracted from the reconstructed birefringence spectrum, and the N1 bias is included in modeling the signal spectrum of the cosmic birefringence. They are, however, negligible compared to the $1\sigma$ statistical error of the reconstructed birefringence spectrum. 

Using a simulation with non-zero birefringence, we confirm that the cross-spectrum between the input and reconstructed birefringence anisotropies agrees with the input spectrum to within $0.3\%$ for multipoles $L\geq 20$, and so we do not apply a transfer function to correct the normalization of the reconstructed spectrum. 

In this paper, we compute the cosmic birefringence power spectrum up to $L\leq 2048$. At larger $L$ values, the statistical uncertainties of the reconstructed spectrum start to increase significantly. The minimum multipole of the reconstructed spectrum is chosen so that the mean-field bias from the global polarization angle uncertainties is negligible (see Sec.~\ref{sec:test}). 

\section{Potential systematics} \label{sec:test}

\begin{figure}[t]
\bc
\includegraphics[width=8.5cm,clip]{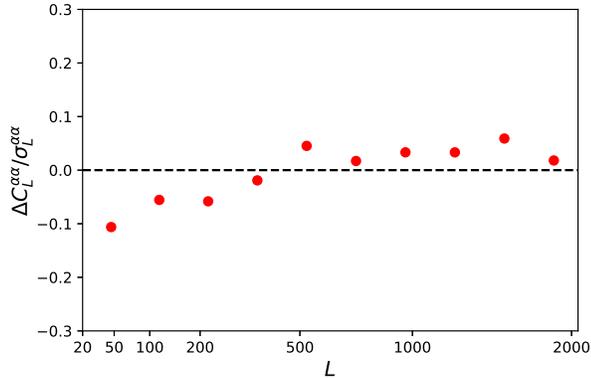}
\caption{
The difference of the cosmic birefringence spectra between the standard plus Galactic dust and standard simulations. Each value has been divided by the $1\,\sigma$ statistical uncertainty in the standard cosmic birefringence spectrum.
}
\label{fig:dust}
\ec
\end{figure}

\begin{figure*}[t]
\bc
\includegraphics[width=8.9cm,clip]{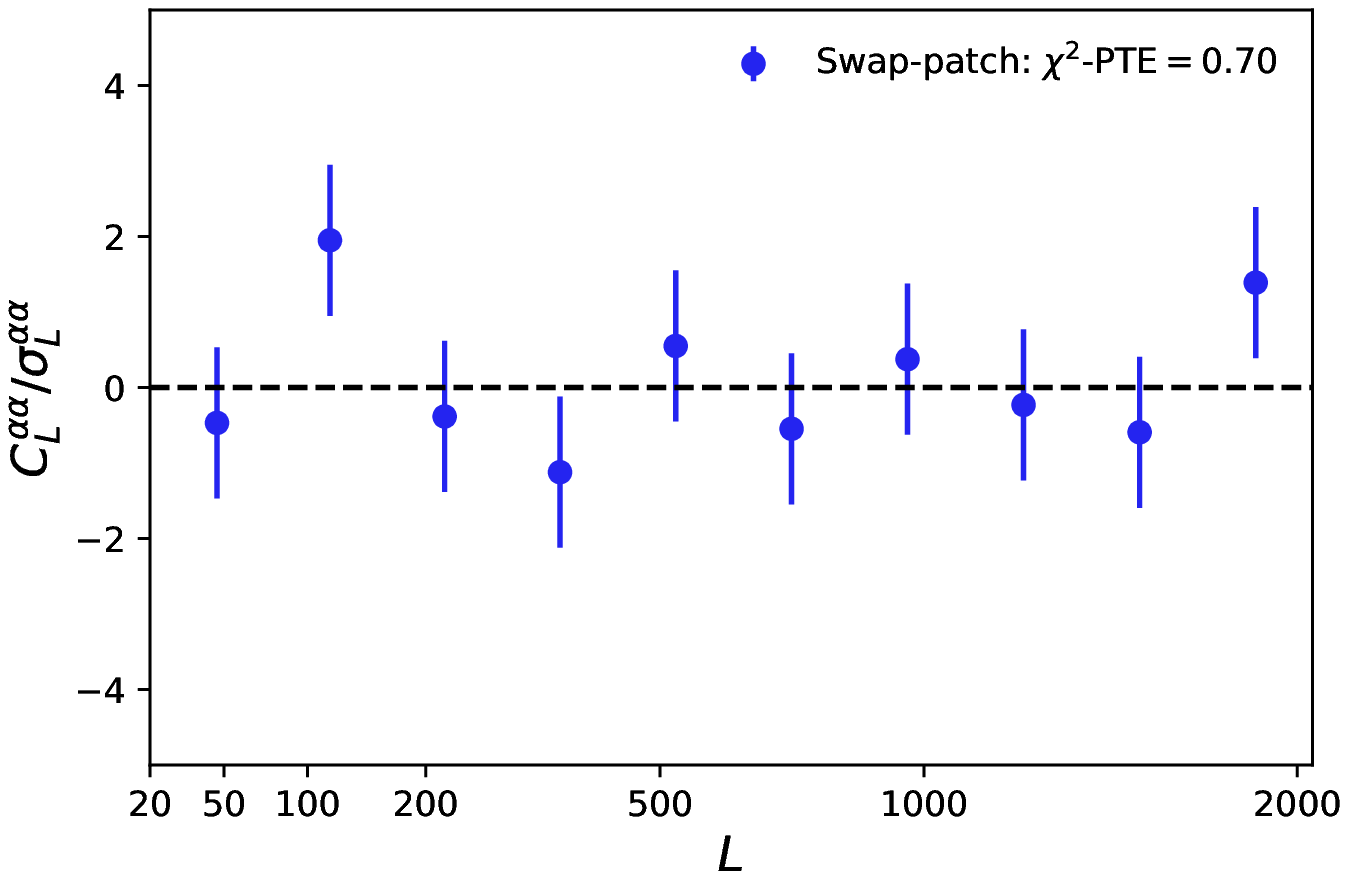}
\includegraphics[width=8.9cm,clip]{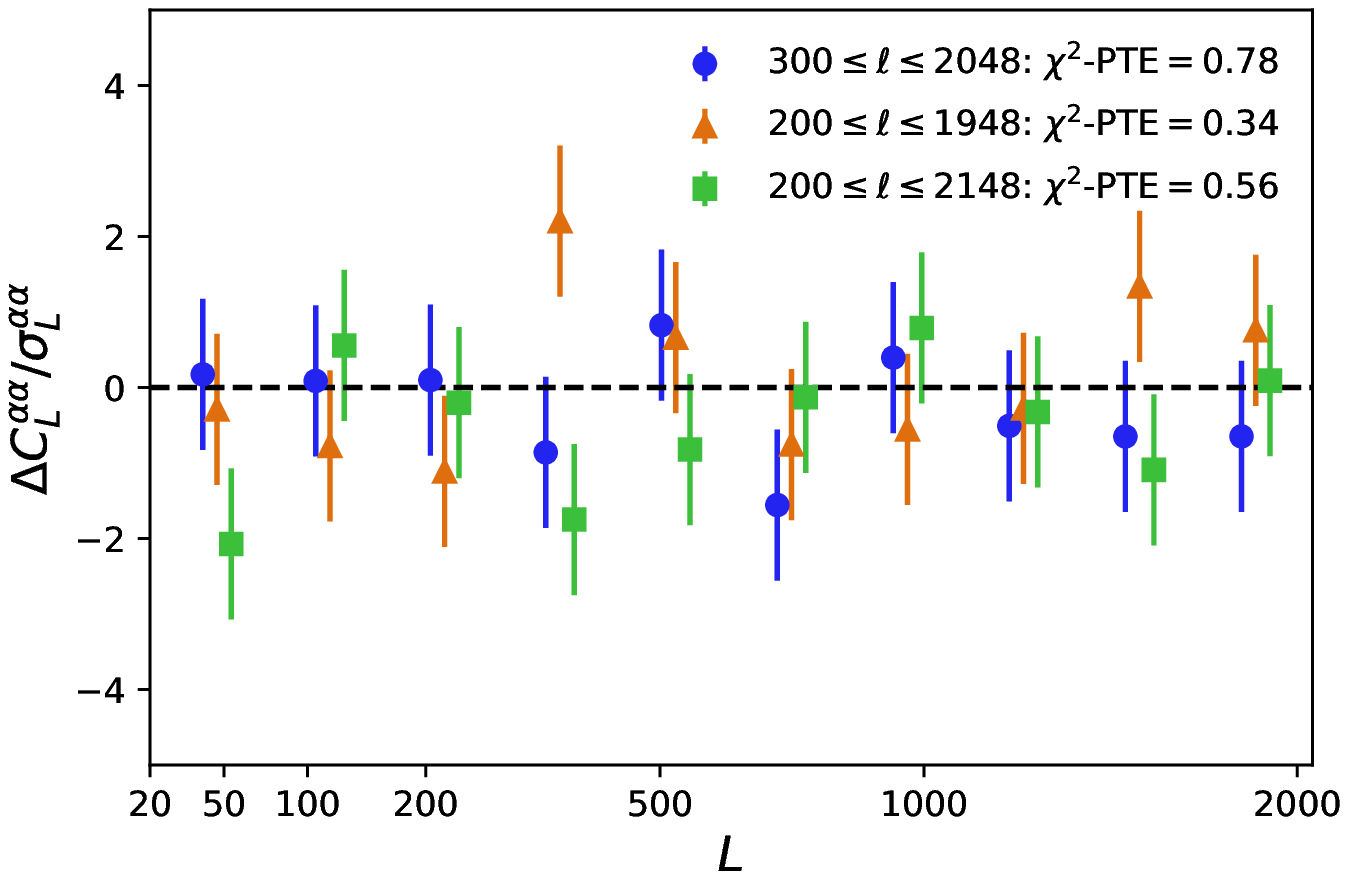}
\caption{
The null cosmic birefringence spectra for the swap patch ({\it left}) and difference spectra ({\it right}) tests, each divided by the statistical $1\sigma$ error of the spectrum. For the swap patch, we show the cross spectrum of the reconstructed cosmic birefringence anisotropies between two separate patches of sky, {\tt D56} and {\tt BOSS-N}.  
}
\label{fig:null}
\ec
\end{figure*}

The ACT polarization data have been tested for possible systematic errors in several published or forthcoming papers focusing on the CMB power spectrum \cite{ACT:Louis:2016}, lensing \cite{ACT16:phi}, and cross-spectra with galaxy surveys \cite{ACT:Omar:2020}. 
Here, we further test for potential systematic contamination which could specifically bias the measured cosmic birefringence spectrum. 
Here and in the following sections, we use $200$ realizations of the simulations to evaluate the band-power covariance matrix for the cosmic birefringence spectrum as well as the chi-squared probability-to-exceed (PTE).

\subsection{Uncertainties in polarization angle measurement}
Global polarization angle errors induce non-zero odd-parity power spectra \cite{Keating:2013,B1rot}. We estimate a constant global rotation angle $\psi$ as follows: Assuming $|\psi|\ll 1$, the global rotation angle is related to the polarization spectra as $\psi=C_b^{EB}/2(C^{EE}_b-C^{BB}_b)\equiv a_b$ at each multipole bin $b$ \cite{Keating:2013}. We compute the angle by minimizing $\sum_{bb'}(\psi-\widehat{a}_b){\rm Cov}^{-1}_{bb'}(\psi-\widehat{a}_{b'})$ where $\widehat{a}_b$ is the observed value of $a_b$ and Cov is the covariance of $a_b$ computed from $200$ realizations of the standard simulation. With the polarization spectra at $\lmin\leq\ell\leq 2048$, we find that $\psi=\gangle$. The $1\,\sigma$ uncertainty of the global rotation angle introduces a significant mean-field bias at very large scales (see Appendix \ref{app:mean} for the origin of the bias). The mean-field bias only becomes close to the $1\,\sigma$ statistical error of the cosmic birefringence spectrum at $L\ll 20$. We therefore exclude the large-scale cosmic birefringence spectrum, $L<20$, from our analysis. Note that this scale roughly corresponds to the fundamental mode determined by our patch size. The measured spectrum below this multipole does not have much information on cosmic birefringence signals. 

An additional possible concern is the variation of polarization angle errors over the field. Variations in relative polarization angles between detectors are calibrated based on optical modeling of the telescope and instrument \cite{Koopman:2016:optics}. In order to obtain variation of polarization angle errors over the field at a significant level, one would require that (i) different detectors have significantly different polarization angle errors; (ii) the relative weights of different detectors vary strongly over the map (since otherwise differential detector angle errors would be absorbed into the mean); and (iii) such an effect is not significantly reduced by averaging from repeated scanning (which it should be, as the main driver of the relative detector weight is the atmospheric loading and thermal environment in the telescope, not which sky direction is being observed). 
Another potential source of the variation of polarization angle in the field is due to our map-making process. In the coordinate transformation from focal plane to horizontal plane for each map obtained from each season, frequency and array, we corrected the coordinate rotation using the angle estimated by $EB$ spectrum, instead of the exact value of the rotation angle, for computational convenience. We checked that the resulting angle correction using this method and with exact coordinate rotation angle agrees with each other, but this process could lead to small angle errors depending on the elevation of our scan.
Since all of these effects are individually unlikely to be large, the likelihood of all these taking place at a significant level is very small, and we therefore neglect such effects. (This is further motivated by the fact that potential upper limits are not degraded by any such systematic, since the systematic is not correlated with a true birefringence signal. As both birefringence and polarization angle error spectra must give strictly positive contributions to the estimated birefringence power spectrum, the presence of such a systematic would, in fact, imply stronger constraints on cosmological birefringence from a data-derived upper limit.)

\subsection{Galactic foregrounds} 
The large scale B-modes are significantly contaminated by Galactic foregrounds, and in principle, non-Gaussian polarized foregrounds could also bias the measured birefringence spectrum. 
We expect minimal direct contamination from the Galactic foregrounds since our analysis removes multipoles below $\l=200$ from the maps before reconstructing the cosmic birefringence anisotropies.
For an accurate estimate of any bias to the cosmic birefringence anisotropies, however, we further test the Galactic foreground contributions to our measurement by adding a simulation of Galactic dust to our standard simulation. In particular, we use 20 different realizations of the Galactic dust simulation provided by \cite{Vansyngel:2016fbn} in the {\tt D56} region for this purpose. We scale the dust polarization maps to our observing frequencies following \cite{P16:dustcls,P18:dust}; we assume a modified blackbody spectrum for dust and use the dust spectral index and temperature of \cite{P18:dust}. 
We then add the scaled polarization maps to $20$ realizations of the input of our standard CMB simulation to produce a set of $20$ simulations including dust. Fig.~\ref{fig:dust} shows the difference spectrum between the simulations including dust and the standard simulations averaged over $20$ realizations. The spectrum is further normalized by the $1\,\sigma$ statistical error of the cosmic birefringence spectrum obtained from $200$ realizations of the standard simulation.
Although we do not yet have sufficient multi-frequency data to fully exclude any impact of Galactic foregrounds, we find that the impact of the dust contribution estimated from our simulations is approximately less than $10\%$ of the $1\,\sigma$ statistical uncertainty at each multipole bin. 

\subsection{Null tests}
As a null test, we compute the cross spectrum of the reconstructed $\a$ obtained from the {\tt D56} and {\tt BOSS-N} fields. The reconstructed cosmic birefringence anisotropies on these two patches should not be correlated, and so the cross-spectrum should be zero. Following the same procedures as applied to the {\tt D56} field, the harmonic coefficients of the cosmic birefringence anisotropies from {\tt BOSS-N} are reconstructed using the curved-sky quadratic estimator described in Sec.~\ref{sec:method} and are then cross-correlated with the birefringence map from {\tt D56}. This null spectrum can serve as a valuable test of whether our error bars are correct. Fig.~\ref{fig:null} shows the cross spectrum; we find that the $\chi^2$ PTE of the cross spectrum is within $2\,\sigma$ range, and the spectrum is consistent with null. 

For additional null tests, we compute the difference between the baseline analysis and cases with different choices of CMB multipole ranges used for the rotation angle reconstruction. Fig.~\ref{fig:null} shows the difference spectra. We calculate the $\chi^2$ PTE for the difference spectra as shown in the figure, finding that the difference spectra are consistent with the null hypothesis irrespective of the choice of the CMB multipole range.

\section{Reconstructed spectrum} \label{sec:results}

\begin{figure*}[t]
\bc
\includegraphics[width=17cm,clip]{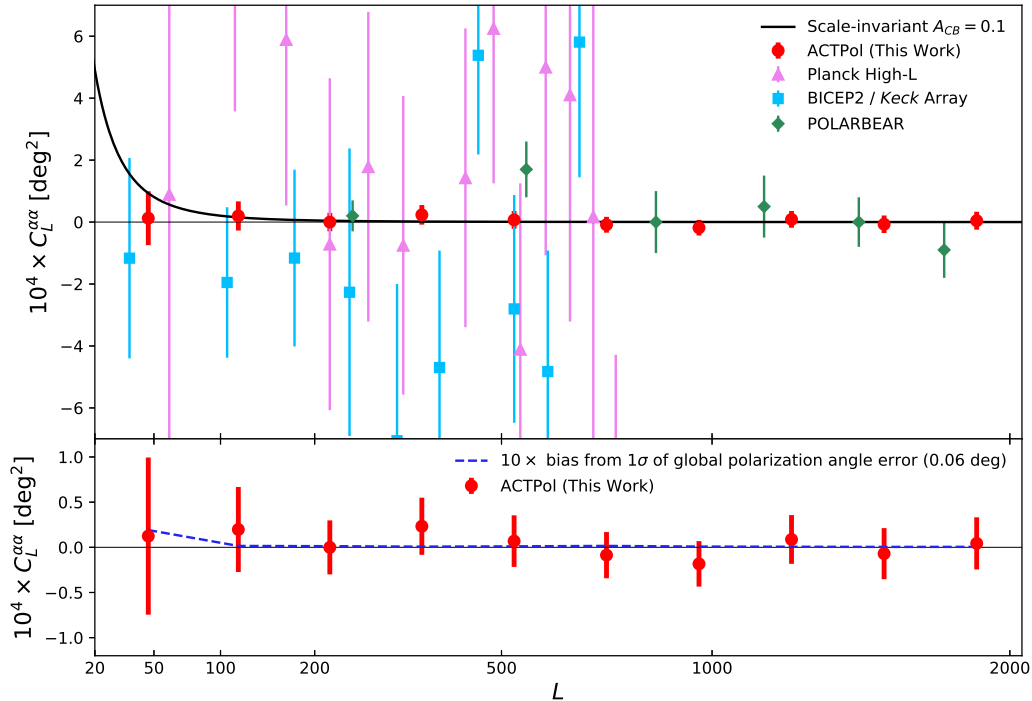}
\caption{
The angular power spectrum of the polarization rotation fields $\alpha(\hatn)$ measured from \actpol\ data over $\areadeep$ deg$^2$ of sky, with errors from a standard $\Lambda$CDM simulation. The solid line shows a scale-invariant spectrum with the amplitude corresponding to our $2\,\sigma$ upper bound (see Sec.~\ref{sec:results}). In addition to our work (red), we also show the spectra obtained from POLARBEAR (green) \cite{PB15:rot}, BICEP2/Keck Array (blue) \cite{BKIX} and Planck (magenta) \cite{Contreras:2017}. The Planck low-$L$ results are not included due to the error bar size. The lower panel shows a zoomed-in view of our birefringence power spectrum measurement; we also show, with a blue dotted line, the potential bias from a global polarization angle systematic error of $0.06$ deg, which is of the same size as the $1\,\sigma$ error from an EB-derived constraint. Since this is difficult to see, for visualization, we have multiplied this angle error bias by a factor of $10$.
}
\label{Fig:aps}
\ec
\end{figure*}

After passing the swap-patch and difference spectrum null tests in Sec.~\ref{sec:test}, we unblinded the reconstructed cosmic birefringence spectrum. Fig.~\ref{Fig:aps} shows the cosmic birefringence spectrum from \actpol\ data with errors obtained from the standard simulation. For comparison, the figure also shows the cosmic birefringence power spectra measured from other recent CMB experiments; the BICEP2/Keck Array \cite{BKIX}, POLARBEAR \cite{PB15:rot}, and Planck \cite{Contreras:2017}. Compared to other experiments, \actpol\ provides the tightest constraint on the cosmic birefringence spectrum at $20\leq L\leq 2048$. We compute the $\chi^2$ PTE of our measured spectrum including covariance obtained from simulation and the value is found to be $\ptereal$; this is in good agreement with zero signal. We note that the off-diagonal elements of the correlation matrix for this measurement become $\sim 0.5$ at $L>1000$ while at lower $L$, off-diagonal band-power correlations are negligible. 

The $\chi^2$ PTE is close to unity; to investigate this further (and test whether this result indicates an overestimate of our error bars), we check the dependency of the $\chi^2$ PTE on analysis choices and summarize in Table~\ref{table:pte}, finding that the value is typically less than $0.95$. Note that the values in Table~\ref{table:pte} are not statistically independent from the baseline value since we only modify the analyzed data by a small amount by changing $L_{\rm min}$ (and changing the number of multipole bins does not introduce any new data). However, if we had significantly overestimated our error bars, we would expect that these other scale ranges and binnings would also have very high PTE values. In addition, as described previously, we have performed several null tests where simulations are used to evaluate the scatter, without finding anomalous PTEs. The high $\chi^2$ PTE for the baseline spectrum therefore is likely due to a statistical fluctuation rather than an overestimate of the errors. Of course, if in fact the errors have been slightly overestimated, our limit on the cosmic birefringence will be somewhat conservative. 

The minimum CMB multipole used in the cosmic birefringence reconstruction is $200$, which is lower than that of the lensing measurement presented in \cite{ACT16:phi}. In that lensing analysis, the CMB multipoles below $\l=500$ are removed since the simulations are not consistent with temperature data at these scales due to inaccurate atmospheric noise characterization and transfer function estimation. For this analysis, however, the temperature data are not used, and the measured polarization noise spectrum is consistent with simulations for $\l \ge 200$. In addition, as demonstrated by our null test in Fig.~\ref{fig:null}, changing the minimum multipole used does not produce any spurious signals. To further test this, we evaluate the $\chi^2$ PTE of a measured spectrum analyzed with $\l_{\rm min}=300$, finding that the value effectively does not change from the case with $\l_{\rm min}=200$; in addition, all our null tests still pass. These facts indicate that the inclusion of low-$\l$ CMB polarization does not introduce non-negligible systematics into our measurement. 

\begin{table}
\bc
\caption{The $\chi^2$ PTE values for our measured cosmic birefringence spectrum with variation of the minimum multipole, $L_{\rm min}$, or the number of multipole bins, $N_b$. For the baseline analysis, where $L_{\rm min}=20$ and $N_b=10$, the PTE is $\ptereal$; the variation seen in this table, given different analysis choices, is consistent with this high PTE being a fluctuation.}
\begin{tabular}{rc|rc}\\
$L_{\rm min}$ & $\chi^2$ PTE & $N_b$ & $\chi^2$ PTE \\\hline\hline
10 & $\pterealLminA$ & 15 & $\pterealbinnA$ \\ 
30 & $\pterealLminB$ & 20 & $\pterealbinnB$ \\ \hline
\end{tabular}
\label{table:pte}
\ec
\end{table}

As an example of the implications of our measurement for phenomenological models of cosmic birefringence, 
we consider a constraint on the amplitude of the scale-invariant spectrum, $C_L^{\a\a}\propto 2\pi/L(L+1)$, which can be later translated into, for example, a constraint on the coupling constant of an axion-like particle. To constrain a scale-invariant spectrum, we first construct an approximate likelihood for the reconstructed cosmic birefringence power spectrum. 
Although we do not use multipoles at $L<20$, the distribution of the power spectrum in the largest bin is asymmetric and is not well described by a Gaussian. Instead, we assume the log-likelihood proposed by \cite{Hamimeche:2008}: 
\al{
    -2\ln \mC{L}(\bm{\hA}) = \sum_{bb'} g(c^0_b\hA_b)[c^1_bC^f_b] {\rm Cov}^{-1}_{bb'}[c^1_{b'}C^f_{b'}]g(c^0_{b'}\hA_{b'})
    \,, \label{Eq:likelihood}
}
where $\hA_b=(\hCaa_b+\ave{\widehat{N}_b^0})/(\Caa_b+\ave{\widehat{N}_b^0})$ is the amplitude of the quadratic-estimator power spectrum relative to that including the cosmic birefringence signals, $\Caa_b$, at each multipole bin, $b$, and $g(x)=\sign(x-1)\sqrt{2(x-\ln x-1)}$ for $x\geq 0$. Note that $\Caa_b$ is the sum of the scale-invariant birefringence spectrum and N1 bias evaluated from the non-zero birefringence simulation. The power spectrum, $C^f_b$, and covariance, ${\rm Cov}_{bb'}$, are evaluated by the mean and variance of the quadratic-estimator power spectrum from the standard simulation, respectively. Note that we further introduce parameters, $c^0_b,c^1_b$, to make the above likelihood closer to that obtained from the simulation. We compute $c^0_b$ and $c^1_b$ by fitting the histogram of $\hA_b$ from the simulation using Eq.~\eqref{Eq:likelihood} at each bin. (We verified that the values of $c^0_b$ and $c^1_b$ only vary by negligible amounts using simulations containing different levels of birefringence signal.)

Using \eq{Eq:likelihood}, we compute the likelihood for the amplitude of the scale-invariant power spectrum defined by $L(L+1)C_L^{\a\a}/2\pi = \ACB\times 10^{-4}$ [rad$^2$]. Assuming a flat prior for $\ACB\geq 0$, we then obtain the $2\,\sigma$ upper limit on the amplitude as $\ACB\leq \ACBconst$ (which is equivalent to $L(L+1)C_L^{\a\a}/2\pi\leq 0.033$ [deg$^2$]). This constraint improves the previous best constraints by a factor of between $2$ and $3$ \cite{Contreras:2017,BKIX}. Note that, for the scale-invariant power spectrum, the constraint on its amplitude is mostly determined by the largest-scale multipole bin; removing the first multipole bin centered at $L=47$ degrades the constraint considerably. 

\section{Discussion} \label{sec:discussion}

Our measured spectrum can be used to constrain various models which lead to cosmic birefringence anisotropies. As an example, we consider the following interaction between axion-like particles and photons in the Lagrangian \cite{Carroll:1989:rot}: 
\al{
    \mS{L} \supset \frac{g_{a\gamma}a}{4}F_{\mu\nu}\tilde{F}^{\mu\nu}
    \,, \label{Eq:gamma-a-int}
}
where $g_{\alpha\gamma}$ is the Chern-Simons coupling constant between the axion-like particles and photon, $a$ is the axion-like particle field, $F_{\mu\nu}$ is the electromagnetic field, and $\tilde{F}^{\mu\nu}$ is its dual. The presence of axion-like particles produces a rotation of the polarization angle as \cite{Carroll:1989:rot,Harari:1992:axion}: 
\al{
	\alpha = \frac{g_{a\gamma}}{2}\Delta a\,, 
}
where $\Delta a$ is the change in $a$ over the photon trajectory. Fluctuations in the axion-like particle field lead to the spatial variation of $\alpha$. 
If the axion-like particle is effectively massless during inflation, the primordial power spectrum of the fluctuations of the axion-like particle field is scale-invariant. As a result, the cosmic birefringence power spectrum becomes a scale-invariant spectrum in the large-scale limit ($L\alt 100$) \cite{Caldwell:2011}:
\al{
	\frac{L(L+1)C_L^{\a\a}}{2\pi} = \left(\frac{H_Ig_{a\gamma}}{4\pi}\right)^2  \,. \label{Eq:scaleinv}
}
Here, $H_I$ is the inflationary Hubble parameter and is related to the tensor-to-scalar ratio, $r$, as $H_I=2\pi M_{\rm pl}\sqrt{A_s r/8}\simeq\sqrt{4r}\times 10^{14}$ GeV where $M_{\rm pl}\simeq 2\times 10^{18}$ GeV is the reduced Planck mass and $A_s\simeq 2\times10^{-9}$ is the amplitude of the primordial scalar perturbations (see, e.g., \cite{Baumann:2009:review}). An axion string network produces a similar scale-invariant spectrum as shown by \cite{Agrawal:2019:biref}. Using \eq{Eq:scaleinv}, our $\ACB$ constraint can be translated into constraints on coupling between axion-like particles and photons as 
\al{
	g_{a\gamma} \leq \frac{\Hgconst}{H_I} = \frac{2.0\times 10^{-16}}{\sqrt{r}} \ \ \text{GeV$^{-1}$}\,. \label{Eq:gag}
}
at $10^{-33}$eV $\alt m_a\alt 10^{-28}$eV.
The coupling constant is related to the decay constant, $f_a$. In string theory models, typically $f_a\sim 10^{16}$ GeV, although it could be the Planck energy scale. Assuming $g_{a\gamma}\sim 10^{-3}/f_a$ \cite{Marsh:2016}, the constraint in \eq{Eq:gag} can be translated into a constraint on $f_a$ as $10^{-6}\alt f_a/H_I$. A detection of the tensor-to-scalar ratio in a future CMB experiment, which would determine $H_I$, would put a lower bound on $f_a$ from the CMB cosmic birefringence. 

The coupling constant of the axion-like particles has been also constrained in various ways and data from astrophysical experiments \cite{Tanabashi:2018:pdg}. The isotropic cosmic birefringence constraint from recent CMB experiments is translated into the constraint on the coupling constant for the axion dark matter as $g_{a\gamma}\alt 1.6\times 10^{-15} (m_a/3\times 10^{-26}{\rm eV})$ GeV$^{-1}$ at $10^{-27}\leq m_a\leq 10^{-24}$ eV \cite{Sigl:2018f:biref}. The interaction described in \eq{Eq:gamma-a-int} causes axion-like particles and photons to interconvert in the presence of a background magnetic field, and the axion-like particles could introduce localized oscillatory modulation in the spectra of photon sources passing through astrophysical magnetic fields. By exploring this effect in the recent X-ray spectral data, \cite{Berg:2016:axion,Reynolds:2019:axion} derive constraint on the coupling constant as $g_{a\gamma}\alt 10^{-12}$ GeV$^{-1}$ for $m_a\alt 10^{-12}$ eV. Coherent oscillations of the Bose condensate of axion-like particles induce periodic changes in the plane of linear polarisation of emission passing through the condensate. Analysis of polarization observations of bright downstream features in the parsec-scale jets of active galaxies leads to a constraint on the coupling constant as $g_{a\gamma}\alt 10^{-12}$ GeV$^{-1}$ for $5\times 10^{-23}\alt m_a\alt 10^{-21}$ eV \cite{Ivanov:2018:axion}. 
Compared to the above other constraints on the lower mass axion-like particles, our constraint on the coupling constant is very stringent if a scenario of large-field inflation models is assumed, $r\agt0.01$. 
To directly compare our constraints with other probes, however, the determination of the energy scale of inflation is necessary. This could be achieved by ongoing and future CMB experiments. 

Our measured spectrum can also be used to constrain other possible sources of the cosmic birefringence proposed by \cite{Liu:2016dcg,Leon:2017,Agrawal:2019:biref}, although constraining these other sources is beyond the scope of this paper.

Measurements of anisotropic cosmic birefringence can be of great importance for testing new physical theories of the early Universe. Future CMB experiments such as the BICEP Array \cite{BICEPArray}, CMB-S4 \cite{CMBS4}, LiteBIRD \cite{LiteBIRD}, Simons Observatory \cite{SimonsObservatory}, and SPT-3G \cite{SPT3G} will measure cosmic birefringence anisotropies even more precisely \cite{Pogosian:2019:axion}. In these experiments, a curved-sky polarization analysis, as we have presented here, will be necessary to tightly constrain a scale-invariant spectrum of cosmic birefringence anisotropies, which is one of the physically best-motivated spectra.


\begin{acknowledgments}
TN thanks Ryo Nagata and Levon Pogosian for helpful discussions. Some of the results in this paper have been derived using a dust simulation generated by \cite{Vansyngel:2016fbn} and public software of the healpy \cite{healpy}, HEALPix \cite{Gorski:2004by}, and CAMB \cite{Lewis:1999bs}. 

This work was supported by the U.S. National Science Foundation through awards AST-1440226, AST0965625 and AST-0408698 for the ACT project, as well as awards PHY-1214379 and PHY-0855887. Funding was also provided by Princeton University, the University of Pennsylvania, and a Canada Foundation for Innovation (CFI) award to UBC. ACT operates in the Parque Astron´omico Atacama in northern Chile under the auspices of the Comisi´on Nacional de Investigaci´on Cient´ıfica y Tecnol´ogica de Chile (CONICYT). Computations were performed on the GPC and Niagara supercomputers at the SciNet HPC Consortium. SciNet is funded by the CFI under the auspices of Compute Canada, the Government of Ontario, the Ontario Research Fund – Research Excellence; and the University of Toronto. The development of multichroic detectors and lenses was supported by NASA grants NNX13AE56G and NNX14AB58G. Colleagues at AstroNorte and RadioSky provide logistical support and keep operations in Chile running smoothly. We also thank the Mishrahi Fund and the Wilkinson Fund for their generous support of the project.

TN, OD and BDS acknowledge support from an Isaac Newton Trust Early Career Grant and from the European Research Council (ERC) under the European Union’s Horizon 2020 research and innovation programme (Grant agreement No. 851274). BDS further acknowledges support from an STFC Ernest Rutherford Fellowship. DH, AM and NS acknowledge support from NSF grant number 1513618. JD and ES acknowledge support from NSF grant number 1814971. EC is supported by an STFC Ernest Rutherford Fellowship ST/M004856/2 and STFC Consolidated Grant ST/S00033X/1. LM received funding from CONICYT FONDECYT grant 3170846. KM acknowledges support from the National Research Foundation of South Africa.

\end{acknowledgments}

\onecolumngrid
\appendix

\section{Separable forms for computing the cosmic birefringence quadratic estimator} \label{app:code}

Here we describe computationally efficient ways for calculating the quadratic estimator of  cosmic birefringence implemented in \url{https://toshiyan.github.io/clpdoc/html/}. 

\subsection{Unnormalized quadratic estimator}

The unnormalized quadratic estimator is given by
\al{
	\uesta_{LM} = \sum_{\l\l'mm'}\Wjm{\l}{\l'}{L}{m}{m'}{M}(-W^-_{\l'L\l})\tCEE_{\l}\ol{E}_{\l m}\ol{B}_{\l'm'}
	\,. \label{Eq:app:uest}
}
Using the properties of the Wigner 3j-symbols and the relationship between the Wigner 3j-symbols and spherical harmonics, we obtain
\al{
    -\Wjm{\l}{\l'}{L}{m}{m'}{M}W^-_{\l'L\l}
    &= [1+(-1)^{\l+\l'+L}]\sqrt{\frac{(2\l+1)(2\l'+1)(2L+1)}{4\pi}}\Wjm{\l'}{L}{\l}{-2}{0}{2}\Wjm{\l}{\l'}{L}{m}{m'}{M}
    \notag \\
    &= \sqrt{\frac{(2\l+1)(2\l'+1)(2L+1)}{4\pi}}\left[\Wjm{\l}{\l'}{L}{2}{-2}{0}+\Wjm{\l}{\l'}{L}{-2}{2}{0}\right]\Wjm{\l}{\l'}{L}{m}{m'}{M}
    \notag \\
    &= \Int{}{\hatn}{}Y_{LM}[Y^{2}_{\l m}Y^{-2}_{\l'm'}+Y^{-2}_{\l m}Y^{2}_{\l'm'}]
    \,.
}
Substituting the above equation into \eq{Eq:app:uest}, we obtain the separable form of the quadratic estimator; 
\al{
	\uesta_{LM} &= \Int{}{\hatn}{}Y^*_{LM}\big[
	\sum_{\l m}Y^2_{\l m}\tCEE_{\l}\ol{E}_{\l m}\sum_{\l'm'}Y^{-2}_{\l'm'}\ol{B}_{\l'm'}+\sum_{\l m}Y^{-2}_{\l m}\tCEE_{\l}\ol{E}_{\l m}\sum_{\l'm'}Y^2_{\l'm'}\ol{B}_{\l'm'}\big]
	\notag \\
	&= \iu \Int{}{\hatn}{}Y^*_{LM}\left[(Q^E+\iu U^E)(Q^B-\iu U^B)-{\rm c.c.}\right]
	\notag \\
	&= -2\Int{}{\hatn}{}Y^*_{LM}\left[U^EQ^B-Q^EU^B\right]
	\,. 
}
where we define the real quantities $Q^E$, $U^E$, $Q^B$, and $U^B$ as
\al{
    Q^E+\iu U^E &= \sum_{\l m}Y^2_{\l m}\tCEE_{\l}\ol{E}_{\l m}
    \,, \notag \\
    Q^B+\iu U^B &= \sum_{\l m}Y^2_{\l m}\iu\ol{B}_{\l m}
    \,.
}

\subsection{Normalization}

The estimator from the previous section must be normalized (see \eq{Eq:aest}). Here we extend the method for lensing and delensing used in \cite{Smith:2010gu} to cosmic birefringence. The inverse of the normalization is given by 
\al{
	\frac{1}{A_L} = \frac{1}{2L+1}\sum_{\l\l'} |W^-_{\l'L\l}|^2 a_\l b_{\l'}
	\,, 
}
where we define $a_\l=1/\hCBB_\l$ and $b_\l=(\tCEE_\l)^2/\hCEE_\l$. Using the relation between the Wigner 3j-symbols and the Wigner d function, we obtain
\al{
	\frac{1}{A_L} 
	&= \pi \sum_{\l\l'}\frac{2\l+1}{4\pi}a_\l\frac{2\l'+1}{4\pi}b_{\l'}
		8\left[\Wjm{\l}{L}{\l'}{-2}{0}{2}^2 + \Wjm{\l}{L}{\l'}{-2}{0}{2}\Wjm{\l}{L}{\l'}{2}{0}{-2}\right]
	\notag \\
	&= \INT{}{\mu}{}{-1}{1} 4\pi \sum_{\l\l'} \frac{2\l+1}{4\pi}a_\l\frac{2\l'+1}{4\pi}b_{\l'}
		(d^\l_{-2,-2}d^L_{00}d^{\l'}_{22} + d^\l_{-2,2}d^L_{00}d^{\l'}_{2,-2})
	\notag \\ 
	&= \INT{}{\mu}{}{-1}{1} 4\pi (\xi_{-2,-2}^a\xi_{22}^b+\xi_{-2,2}^a\xi_{2,-2}^b) d^L_{00}
	\,, 
}
where we define
\al{
	\xi^a_{mm'} &= \sum_\l\frac{2\l+1}{4\pi}a_\l d^\l_{mm'} 
	\,. 
}

\section{Mean-field bias in the presence of a global polarization angle error} \label{app:mean}

Here we describe how the global polarization angle error introduces the mean-field bias in the rotation estimator. Assuming non-zero $\CEB_\l$, the off-diagonal elements of the $EB$ correlation induced by lensing have the following additional term containing $\CEB_\l$: 
\al{
    \ave{\ol{E}_{\l m}\ol{B}_{\l'm'}}_{\rm CMB} = \sum_{LM}\Wjm{\l}{\l'}{L}{m}{m'}{M} \widetilde{f}^\grad_{\l L \l'} \grad^*_{LM} 
    \,, 
}
where $\widetilde{f}^\grad_{\l L\l'}$ is the usual weight function of the $EE$ quadratic estimator for lensing \cite{OkamotoHu:quad} but replacing the $EE$ with the $EB$ spectrum. For the finite sky coverage, the window function also produces off-diagonal correlations as
\al{
    \ave{\ol{E}_{\l m}\ol{B}_{\l'm'}}_{\rm CMB} = \sum_{LM}\Wjm{\l}{\l'}{L}{m}{m'}{M} \widetilde{f}^\epsilon_{\l L \l'} \epsilon^*_{LM} 
    \,, 
}
where $\widetilde{f}^\epsilon_{\l L\l'}$ is the usual weight function of the $EE$ quadratic estimator for the window \cite{Namikawa:2012:bhe}, but replacing the $EE$ with the $EB$ spectrum. The above equations indicate that, if $\CEB_{\l}$ is non-zero due to a global polarization error, the lensing and window introduces the following mean-field bias: 
\al{
    \ave{\widehat{\alpha}_{LM}} = \frac{A_L}{2L+1}\sum_{x=\grad,\epsilon} x_{LM}\sum_{\l\l'}\frac{f^\alpha_{\l L\l'}\widetilde{f}^x_{\l L\l'}}{\hCEE_\l\hCBB_{\l'}} 
    \,. 
}
We can construct a bias-hardened estimator for $\widehat{\alpha}$ in a similar way as \cite{Namikawa:2012:bhe}. However, the signal-to-noise ratio of the bias-hardened estimator is significantly degraded. 

\twocolumngrid

\bibliographystyle{mybst}
\bibliography{cite}

\end{document}